\documentclass[12pt]{JHEP3}

\preprint{  }

\usepackage{epsfig,multicol}
\usepackage{amsmath}
\usepackage{graphics}
\usepackage{graphicx}
\usepackage{dcolumn}
\usepackage{amssymb}
\usepackage{amsthm}
\usepackage{amsfonts}
\usepackage{subfigure}
\usepackage{setspace}

\title{Geometric phase of two-level atoms and thermal nature of de Sitter spacetime}
\author{Zehua Tian and
Jiliang  Jing\footnote{Corresponding author.}
\\ Department of Physics, and Key Laboratory of Low Dimensional Quantum Structures and Quantum Control of Ministry of Education, Hunan Normal University, Changsha, Hunan 410081, P. R. China
\\
\\
E-mail: 136421865@qq.com, jljing@hunnu.edu.cn}

\abstract{

In the framework of open quantum systems, we study the geometric phase acquired by freely falling and static two-level atoms interacting with quantized conformally coupled massless scalar fields in de Sitter-invariant vacuum. We find that, for the freely falling atom, the geometric phase gets a correction resulting from a thermal bath with the Gibbons-Hawking temperature, thus it clearly reveals the intrinsic thermal nature of de Sitter spacetime from a different physical context. For the static atom, there is a correction to the geometric phase coming from both the intrinsic thermal nature of de Sitter spacetime and the Unruh effect associated with the proper acceleration of the atom. Furthermore, in a gedanken experiment, we estimate the magnitude of the correction to the geometric phase as opposed to that in a flat spacetime. We find that the correction for the freely falling atom is too tiny to be measured, and that for the static atom achieves an observable magnitude only when the atom almost locates at the horizon.
}


\begin{document}

\section{Introduction}

In 1956, Pancharatnam \cite{Pancharatnam} first introduced the concept of geometric phase when studying the interference of classical light in distinct states of polarization. After that, in 1984, Berry \cite{berry} discovered that the state of a quantum system, in addition to the usual dynamical phase, can acquire a purely geometric phase after a cyclic adiabatic evolution. Because this phase provides us a very deep insight into the geometric structure of quantum mechanics and gives rise to various observable effects, a lot of efforts have been taken to generalize Berry's work to different evolutions, such as nonadiabatric evolution \cite{Aharonov} and noncyclic evolution \cite{Samuel}. Furthermore, the geometric phase, besides theoretical studies, also has been studied in practical experiments. For example, by using the nuclear magnetic resonance technique the mixed state geometric phase \cite{Sjoqvist 1} has been verified experimentally in Ref. \cite{Du}.

 All real world quantum systems, as we all known, inevitably interact with surrounding environment. So, studying the geometric phase of open quantum system, which can make our theoretical models be closer to really physical process, is very significant and attracts much interest recently \cite{Fujikawa,Huang}. However, when the study relates to an open quantum system, many physical concepts, such as nonunitary evolution and mixed state, have to be introduced to exactly and completely describe the evolution of the open quantum system. Because of that, it is very important to extent the geometric phase to mixed state and nonunitary evolution cases. In this regard, let's note that Uhlmann \cite{Uhlmann} first defined a mixed-state geometric phase via mathematical concept of purification. Sj\"{o}qvist \emph{et~al.} \cite{Sjoqvist 1} introduced an alternative definition of geometric phase for nondegenerate density operators based upon quantum interferometry. The extension to degenerate mixed state has soon been done by Singh \emph{et~al.} \cite{Singh}. For nonunitary evolution case, these generalizations has been shown in Refs. \cite{Ericsson,Faria}. However, the concept proposed in Ref. \cite{Ericsson}, when using different Kraus representations, may yield different values of geometric phase for a given completely positive map. Defects also exist in Ref. \cite{Faria}, whose results may not reduce to the expected results \cite{Sjoqvist 1,Singh} in the limit of unitary evolution due to a weaker form of parallel transport condition than Ref. \cite{Ericsson} used. Considering this, in Ref. \cite{Tong}, D. M. Tong \emph{et~al.}, from kinematic approach, gave an expression of the mixed-state geometric phase in the nonunitary evolution, this phase is manifestly gauge invariant and can be experimentally tested in interferometry. After that, the study of geometric phase in open quantum system became more extensive
 \cite{Fujikawa,Huang,Hu 1,Hu 2}.

Recently, the combination of geometric phase and relativistic effect has been widely studied. Martin-Martinez \emph{et~al.} \cite{Martin} theoretically considered  the possibility of using geometric phase to detect the Unruh effect. J. Hu \emph{et~al.} \cite{Hu 1}, by using open quantum system approach, analysed geometric phase for an accelerated two-level atom. Not only that, since any systems, in quantum sense, will be subject to vacuum fluctuations, they also generalized the geometric phase, which is acquired by a two-level atom coupling to vacuum fluctuations, to the background of Schwarzschild black hole \cite{Hu 2}. Because of this coupling, one naturally expect that some physical properties of vacuum will be reflected in the observable phenomena of quantum system, such as Lamb shift \cite{zhou1,zhou2,Rizzuto} and geometric phase, when this system evolves in the vacuum.

In this paper, we will study the geometric phase of both freely falling and static atoms interacting with quantized conformally coupled massless scalar fields in the de Sitter-invariant vacuum.
The reason for special attention to de Sitter spacetime in recent years stems from the fact that de Sitter space is the unique maximally symmetric curved spacetime. It enjoys the same degree of symmetry as Minkowski space (ten Killing vectors). Besides, Our current observations, together with the theory of inflation, suggest that our universe may approach de Sitter geometries in the far past and the far future. And a duality may exist between quantum gravity on de Sitter spacetime and a conformal field theory living on the boundary identified with the timelike infinity of it \cite{Strominger}. So, many fields, such as fields quantization \cite{Mottola,Allen,Allen1,Bunch,Mishima,Polarski1,Polarski2,Nakayama,Polarski3,Galtsov}, Lamb shift \cite{zhou1} and spontaneous excitation of atom \cite{Yu}, have been studied in this special curved spacetime, and so we will focus our attention on this spacetime to study the geometric phase of atom.

It is needed to note that we use open quantum system approach to study the geometric phase,
this approach is different from that adopted by Martin-Martinez \emph{et~al.} in Ref. \cite{Martin}. They modeled the detector by a harmonic oscillator which couples only to a single mode of scalar field. They diagonalized exactly the total Hamiltonian (atom+field), used the unitary operator to evolve the state and calculated the phase. Because of the single-mode interaction, a cavity must be introduced. And the cavity, in order to avoid that the boundary conditions inhibit the Unruh effect, is assumed to be transparent to the field mode to which the detector couples. Such kind of cavity seems to be a major challenge in experimental implementation of their proposal. However, in our paper, we consider that a two-level system, which is treated as an open system in a reservoir of fluctuating vacuum scalar field, couples to all modes of this field. We obtain the evolved state from open quantum system approach, and calculate the geometric phase of it by using the generalized definition of geometric phase for nonunitary evolutions. Since in our model, the atom couples to all vacuum modes of field, and no cavity is needed in any experimental scheme, our approach, in this regard, is better than that used in Ref. \cite{Martin}. Furthermore, the quantum geometric phase of an open system undergoing nonunitary evolution due to its coupling to a quantum critical bath has recently been demonstrated using a NMR quantum simulator \cite{critical}.

Our paper is constructed as follows: after briefly reviewing quantum evolution of a two-level atom and simply calculating the geometric phase of evaluated atomic state in section \ref{section 1}, we calculate and discuss the geometric phase of freely falling atom in de Sitter spacetime in section \ref{section 2} and that of static atom case in section \ref{section 3}. In section \ref{section 4} we give our conclusions.

\section{Quantum evolution of a two-level atom and geometric phase} \label{section 1}

Let's begin with the Hamiltonian of the system containing atom and external field, which is given by
\begin{eqnarray}\label{Hamiltonian}
H=H_s+H_\phi+H_I,
\end{eqnarray}
where $H_s$ and $H_\phi$ are the Hamiltonian of atom and scalar field, respectively, and $H_I$ represents their interaction. For simplicity, we take a two-level atom with Hamiltonian $H_s=\frac{1}{2}\omega_0\sigma_z$, where $\omega_0$ is the energy level spacing of the atom, and $\sigma_z$ is the Pauli matrix. We assume that the Hamiltonian describing the interaction between atom and scalar field is $H_I=\mu(\sigma_++\sigma_-)\phi(x(\tau))$, in which $\sigma_+$ ($\sigma_-$) is the atomic rasing (lowering) operator, and $\phi(x)$ corresponds to the scalar field operator in de Sitter spacetime.

Initially the total density operator of the system plus field is assumed to be $\rho_{tot}=\rho(0)\otimes|0\rangle\langle0|$, in which $\rho(0)$ is the reduced density matrix of the atom, and $|0\rangle$ is the vacuum of the field. For the total system, its equation of motion is given by
\begin{eqnarray}\label{motion equation}
\frac{\partial\rho_{tot}(\tau)}{\partial\tau}=-i[H,\rho_{tot}(\tau)],
\end{eqnarray}
where $\tau$ is the proper time of the atom. In the limit of weak coupling, the evolution of the reduced density matrix $\rho(\tau)$, after simplification, can be written in the Lindblad form \cite{Lindblad,Benatti}
\begin{eqnarray}\label{Lindblad equation}
\frac{\partial\rho(\tau)}{\partial\tau}=-i[H_{eff},\rho(\tau)]+\cal L[\rho(\tau)],
\end{eqnarray}
with
\begin{eqnarray}\label{Lindblad operator}
{\cal L}[\rho]=\sum^3_{j=1}[2L_j\rho L^\dagger_j-L^\dagger_jL_j\rho-\rho L^\dagger_jL_j],
\end{eqnarray}
where $H_{eff}$ and $L_j$ relate to the Pauli matrices and the field correlation function
\begin{eqnarray}\label{correlation function}
G^+(x-x')=\langle0|\phi(x)\phi(x')|0\rangle.
\end{eqnarray}
We define
\begin{eqnarray}
\nonumber
\gamma_\pm&=&4\mu^2\mathrm{Re}\Gamma_\pm=2\mu^2\int^{+\infty}_{-\infty}e^{\mp i\omega_0s}G^+(s-i\epsilon)ds,\nonumber\\
\gamma_z&=&0, \end{eqnarray}
with $s=\tau-\tau'$, then we have
\begin{eqnarray}\label{L values}
\nonumber
H_{eff}=\frac{1}{2}\Omega\sigma_z=\frac{1}{2}\{\omega_0+2\mu^2\mathrm{Im}(\Gamma_++\Gamma_-)\}\sigma_z,
\\
L_1=\sqrt{\frac{\gamma_-}{2}}\sigma_-,~~L_2=\sqrt{\frac{\gamma_+}{2}}\sigma_+,
~~L_3=\sqrt{\frac{\gamma_z}{2}}\sigma_z.
\end{eqnarray}
For convenience, we write the density matrix $\rho(\tau)$ in terms of the Pauli matrices
\begin{eqnarray}\label{density matrices}
\rho(\tau)=\frac{1}{2}\left(1+\sum^3_{i=1}\rho_i(\tau)\sigma_i\right).
\end{eqnarray}
By substituting Eq. (\ref{density matrices}) into Eq. (\ref{Lindblad equation}), and assuming the initial state of the atom $|\psi(0)\rangle=\cos\frac{\theta}{2}|1\rangle+\sin\frac{\theta}{2}|0\rangle$, after some calculations, we finally get the time-dependent reduced density matrix
\begin{eqnarray}\label{time-dependent reduced density}
\rho(\tau)=\frac{1}{2}\left(
\begin{array}{cc}
1+e^{-A\tau}\cos\theta+\frac{B}{A}(1-e^{-A\tau}) & e^{-\frac{1}{2}A\tau-i\Omega\tau}\sin\theta \\
e^{-\frac{1}{2}A\tau+i\Omega\tau}\sin\theta & 1-e^{-A\tau}\cos\theta-\frac{B}{A}(1-e^{-A\tau})
\end{array}
\right),
\end{eqnarray}
where $A=\gamma_++\gamma_-$ and $B=\gamma_+-\gamma_-$.

As introduced in \cite{Tong}, the geometric phase of a mixed state undergoing nonunitary evolution is given by
\begin{eqnarray}\label{geometric phase}
\Phi=\mathrm{arg}\left(\sum^N_{k=1}\sqrt{\lambda_k(0)\lambda_k(T)}\langle\phi_k(0)|\phi_k(T)\rangle
e^{-\int^T_0\langle\phi_k(\tau)|\dot{\phi}_k(\tau)\rangle d\tau}\right),
\end{eqnarray}
where $\lambda_k(\tau)$ and $|\phi_k(\tau)\rangle$ are respective the eigenvalues and eigenvectors of the reduced density matrix $\rho(\tau)$. Obviously, to calculate the geometric phase of a state, one must first get its eigenvalues. For state (\ref{time-dependent reduced density}), its eigenvalues are
\begin{eqnarray}\label{eigenvalues}
\lambda_\pm(\tau)=\frac{1}{2}(1\pm\eta),
\end{eqnarray}
where $\eta=\sqrt{\rho^2_3+e^{-A\tau}\sin^2\theta}$ with $\rho_3=e^{-A\tau}\cos\theta+\frac{B}{A}(1-e^{-A\tau})$.
It is interesting to note that only the eigenvector corresponding to $\lambda_+(\tau)$ has the contribution to geometric phase because $\lambda_-(0)=0$. Thus, we just give the eigenvector corresponding
to $\lambda_+(\tau)$
\begin{eqnarray}\label{eigenvector}
|\phi_+(\tau)\rangle=\sin\frac{\theta_\tau}{2}|1\rangle +\cos\frac{\theta_\tau}{2}e^{i\Omega\tau}|0\rangle,
\end{eqnarray}
with
\begin{eqnarray}
\tan\frac{\theta_\tau}{2}=\sqrt{\frac{\eta+\rho_3}{\eta-\rho_3}}.
\end{eqnarray}
Then, according to Eq. (\ref{geometric phase}), we obtain the geometric phase for an time interval $T$ of evolution,
\begin{eqnarray}\label{phase}
\nonumber
\Phi&=&-\Omega\int^T_0\cos^2\frac{\theta_\tau}{2}d\tau
\\
&=&-\int^T_0\frac{1}{2}\left(1-\frac{Re^{A\tau}-R+\cos\theta}
{\sqrt{e^{A\tau}\sin^2\theta+(Re^{A\tau}-R+\cos\theta)^2}}\right)\Omega d\tau,
\end{eqnarray}
where $R=\frac{B}{A}$. After integral, we obtain
\begin{eqnarray}\label{final phase}
\Phi=P[\varphi]-P[0],
\end{eqnarray}
with
\begin{eqnarray}\label{phase parameter}
\nonumber
P[\varphi]&=&-\frac{\Omega}{2\omega_0}\big\{\varphi-\frac{\omega_0}{A}\frac{R}{|R|}
\ln\big(\frac{1-Q^2-R^2+2R^2e^{A\varphi/\omega_0}}{2|R|}+S(\varphi)\big)
\\
&&-\frac{\omega_0}{A}\frac{Q}{|Q|}\ln\left(1-Q^2-R^2+2Q^2e^{-A\varphi/\omega_0}
+2|Q|S(\varphi)e^{-A\varphi/\omega_0}\right)\big\}, \end{eqnarray}
where $S(\varphi)=\sqrt{R^2e^{2A\varphi/\omega_0}+(1-Q^2-R^2)e^{A\varphi/\omega_0}+Q^2}$, $\varphi=\omega_0T$ and $Q=R-\cos\theta$.

The expression (\ref{phase}) is a general expression of the geometric phase. For different evolutions, we can obtain different geometric phase because the parameters $A$ and $B$ have different values. In the following, we will calculate the geometric phase for two special cases, one is a freely falling atom in de Sitter spacetime, and the other is a static atom in de Sitter spacetime.


\section{Geometric phase of a freely falling atom in de Sitter spacetime} \label{section 2}

In this section, we consider the geometric phase of a freely falling atom interacting with
a quantized conformally coupled massless scalar field in de Sitter spacetime. As shown in \cite{Mottola,Birrell}, several different coordinate systems can be chosen to characterize de Sitter spacetime. Here we choose the global coordinate system $(t,\chi,\theta,\phi)$ under which the freely falling atom is comoving with the expansion, and the corresponding line element of which is given by
\begin{eqnarray}\label{line element 1}
ds^2=dt^2-\alpha^2\cosh^2(t/\alpha)[d\chi^2+\sin\chi^2 (d\theta^2+\sin\theta^2d\phi^2)],
\end{eqnarray}
with $\alpha=\sqrt{\frac{3}{\Lambda}}$, where $\Lambda$ is the cosmological constant. If
$-\infty<t<\infty$, $0\leq\chi\leq\pi$, $0\leq\theta\leq\pi$, $0\leq\phi\leq2\pi$, then
the coordinate covers the whole de Sitter manifold \cite{Mottola,Allen,Birrell}.

It is worth noting that the canonical quantization of the scalar field with the above metric has been done widely \cite{Allen,Allen1,Bunch,Mishima}, and for the massless scalar field the Wightman function for
the freely falling atom, in the conformally coupling limit, can be represented as
\begin{eqnarray}\label{Wightman function 1}
G^+(x-x')=-\frac{1}{16\pi^2\alpha^2\sinh^2(\frac{\tau-\tau'}{2\alpha}-i\epsilon)}.
\end{eqnarray}
For this Wightman function, $A$, $B$ and $H_{eff}$ introduced above are given by
\begin{eqnarray}
A&=&\frac{\mu^2\omega_0}{\pi}\left(\frac{e^{2\pi\alpha\omega_0}+1} {e^{2\pi\alpha\omega_0}-1}\right),
\nonumber \\ B&=&-\frac{\mu^2\omega_0}{\pi}, \nonumber \\
H_{eff}&=&\frac{1}{2}\{\omega_0+2\mu^2\mathrm{Im}(\Gamma_++\Gamma_-)\}\sigma_z
\nonumber \\
&=&\frac{1}{2}\big\{\omega_0+\frac{\mu^2}{2\pi^2}\int^\infty_0d\omega
P(\frac{\omega}{\omega+\omega_0}-\frac{\omega}{\omega-\omega_0})(1+\frac{2}{e^{2\pi\alpha\omega}-1})\big\}
\sigma_z, \label{Lam shift} \end{eqnarray}
where the last term of $H_{eff}$ represents the Lamb shift of the freely falling atom in de Sitter spacetime. Here we will simply discuss the Lamb shift of the atom although which is not our focus in this paper. We rewrite the term of Lamb shift in Eq. (\ref{Lam shift}) as
\begin{eqnarray}\label{Lamb shift term}
\nonumber
\Delta&=&\Delta_0+\Delta_{T_f},
\end{eqnarray}
with
\begin{eqnarray}
\Delta_0&=&\frac{\mu^2}{4\pi^2}\int^\infty_0d\omega
P(\frac{\omega}{\omega+\omega_0}-\frac{\omega}{\omega-\omega_0}),
\nonumber \\
\Delta_{T_f}&=&\frac{\mu^2}{2\pi^2}\int^\infty_0d\omega
P(\frac{\omega}{\omega+\omega_0}-\frac{\omega}{\omega-\omega_0}) \frac{1}{e^{2\pi\alpha\omega}-1},
\end{eqnarray}
where $\Delta_0$ is just the Lamb shift of an inertial two-level atom in a free Minkowski spacetime. Obviously, it is logarithmically divergent, but this divergence can be removed by introducing a cutoff on the upper limit of the integral. $\Delta_{T_f}$ results from the thermal effect of de Sitter spacetime felt by the freely falling atom. The evaluation of this
integral must be done numerically, which is a finite value due to the tiny Gibbons-Hawking temperature. The similar discussions about Lamb shift of an accelerated atom in Minkowski spacetime have been done in Ref. \cite{Audretsch}. It is worth noting that the same result of the Lamb shift in de Sitter spacetime has been obtained by using Dalibard, Dupont-Roc and Cohen-Tannoudji formalism \cite{zhou1}.

In Eq. (\ref{Lam shift}), the Lamb shift is very weak compared with $\omega_0$, thus we can omit the Lamb shift term when calculating
the geometric phase of the atom. For small $\frac{\gamma_0}{\omega_0}=\frac{\mu^2}{\pi}$,
where $\gamma_0=\frac{\mu^2\omega_0}{\pi}$ is the spontaneous emission rate, we may Taylor expand the
geometric phase and obtain, to the first order of $\frac{\gamma_0}{\omega_0}$,
\begin{eqnarray}\label{GP for freely falling atom}
\Phi_f=-\varphi\sin^2\frac{\theta}{2}-\varphi^2\frac{\gamma_0}{8\omega_0}\sin^2\theta
(2+\cos\theta+\frac{2}{e^{2\pi\alpha\omega_0}-1}\cos\theta).
\end{eqnarray}
The first term $-\varphi\sin^2\frac{\theta}{2}$ in Eq. (\ref{GP for freely falling atom}) denotes the geometric
phase of an isolated two-level atom, and the second term is the correction due to the interaction between the freely falling atom with environment. Furthermore, in the limit of $\alpha\rightarrow\infty$, Eq. (\ref{GP for freely falling atom}) corresponds to the case of an inertial atom in Minkowski spacetime. Obviously, a correction still exists, which is the consequence of the zero point fluctuations of the Minkowski vacuum. The explicit form of this correction reads
\begin{eqnarray}\label{geometric phase inertial atom}
\Phi_I=\lim_{\alpha\rightarrow\infty} \Phi_f =-\varphi\sin^2\frac{\theta}{2}-\varphi^2\frac{\gamma_0}{8\omega_0}\sin^2\theta
(2+\cos\theta).
\end{eqnarray}
In order to obtain the correction to the geometric phase purely resulting from the effect of de Sitter spacetime, the contribution deriving from the inertial atom in Minkowski spacetime needs to be got rid of from
Eq. (\ref{GP for freely falling atom}). After doing that, we obtain
\begin{eqnarray}\label{pure geometric phase 1}
\delta\Phi=\Phi_f-\Phi_I=-\varphi^2\frac{\gamma_0}{4\omega_0}\sin^2\theta \frac{1}{e^{2\pi\alpha\omega_0}-1}\cos\theta.
\end{eqnarray}
Obviously, a thermal factor $(e^{2\pi\alpha\omega_0}-1)^{-1}$ appears here. This form is similar to the correction to that of an inertial atom immersed in a thermal bath in a Minkowski spacetime at the temperature $T_f=1/2\pi\alpha$. Therefore, for a freely falling atom in de Sitter spacetime, a thermal-like term revise exists for the geometric phase as opposed to that in the Minkowski sapcetime, and this correction relates to a temperature $T_f=1/2\pi\alpha$, which is exactly the Gibbons-Hawking temperature \cite{Hawking}.

It is believed that our universe will approach to a de Sitter space in the future although it is expanding in current. So, we may ask whether the correction to the geometric phase given here can be detected. To answer this question, in the following, we will estimate the magnitude of this correction. From Eq. (\ref{pure geometric phase 1}), we can read that the correction depends on the translation frequency $\omega_0$ and the spontaneous emission rate $\gamma_0$ of the atom, the initial state characterized by $\theta$, the parameterized cosmological constant $\alpha$, and $\varphi$ relating to the time interval of evolution. According to current observations, it is estimated that the Gibbons-Hawking temperature is of only $\sim3.94\times10^{-30}~\mathrm{K}$. If we choose physically relevant frequency of atom transitions $\omega_0\simeq1.89~\mathrm{GHz}$, then the coefficient $e^{\hbar\omega_0/(\kappa_BT_f)}$ is of the order of $\sim10^{10^{28}}$, which is a extremely large value. So, although we, since the geometric phase accumulates, can enhance the phase difference by evolving the system through more cycles and obtain $\varphi^2\sim5.74\times10^{32}$ for one year, the phase difference $\delta\Phi$ is still extremely tiny. Therefore, the correction is unrealistic for actual experimental measurement.


\section{Geometric phase of a static atom in de Sitter spacetime}\label{section 3}

Next we will calculate the geometric phase of a static atom in interaction
with a conformally coupled massless scalar field in de Sitter-invariant vacuum. For this purpose,
the static coordinate system will be chosen, and corresponding line element is given by
\begin{eqnarray}\label{static coordinate}
ds^2=\big(1-\frac{r^2}{\alpha^2}\big)d\widetilde{t}^2-\big(1-\frac{r^2}{\alpha^2}\big)^{-1}dr^2
-r^2(d\theta^2+\sin^2\theta d\phi^2).
\end{eqnarray}
We can see from Eq. (\ref{static coordinate}) that this metric possesses a event horizon at $r=\alpha$, and
usually we call it cosmological horizon. Note that the coordinates $(\widetilde{t},r,\theta,\phi)$ only cover part of de Sitter spacetime, just like the Rindler wedge. An atom at rest in this static coordinates system has the proper acceleration
\begin{eqnarray}\label{proper acceleraion}
a=\frac{r}{\alpha^2}\big(1-\frac{r^2}{\alpha^2}\big)^{-1/2}.
\end{eqnarray}
And the relation between the static and global coordinates system is
\begin{eqnarray}\label{static and global relation}
r=\alpha\cosh(t/\alpha)\sin\chi,~~~~\tanh(\widetilde{t}/\alpha)=\tanh(t/\alpha)\sec\chi.
\end{eqnarray}
It is of interest to note that the worldline $r=0$ in the static coordinate coincides with the worldline $\chi=0$ in the global coordinate, and an atom at rest with $r\neq0$ in the static coordinate will be accelerated relative to the observer at rest in the global coordinate with $\chi=0$.

By solving the field equation in the static coordinates system, a set of modes will be obtained \cite{Mishima,Polarski1,Polarski2,Nakayama}. In a de Sitter-invariant vacuum, we can calculate the Wightman function for the massless conformally coupled scalar field, and which is given by \cite{Polarski3,Galtsov}
\begin{eqnarray}\label{wightman function 2}
G^+(x-x')=-\frac{1}{8\pi^2\alpha^2}\frac{\cosh(\frac{r^*}{\alpha})\cosh(\frac{{r^*}'}{\alpha})}
{\cosh(\frac{\widetilde{t}-{\widetilde{t}}'}{\alpha}-i\epsilon)
-\cosh(\frac{r^*-{r^*}'}{\alpha})},
\end{eqnarray}
with $r^*=\frac{\alpha}{2}\ln\frac{\alpha+r}{\alpha-r}$. So, for a static atom, it can be simplified to
\begin{eqnarray}\label{wightman function 3}
G^+(x-x')=-\frac{1}{16\pi^2\kappa^2\sinh(\frac{\tau-\tau'}{2\kappa}-i\epsilon)},
\end{eqnarray}
where $\kappa=\sqrt{g_{00}}\alpha$ and $\tau=\sqrt{g_{00}}\widetilde{t}$. Comparing the Wightman function (\ref{wightman function 3}) with (\ref{Wightman function 1}), it is easy to obtain
\begin{eqnarray}
A=\frac{\mu^2\omega_0}{\pi}\left(\frac{e^{2\pi\kappa\omega_0}+1}{e^{2\pi\kappa\omega_0}-1}\right),
~~~~~B=-\frac{\mu^2\omega_0}{\pi}, \end{eqnarray}
and
\begin{eqnarray}\label{Lam shift1}
\nonumber
H_{eff}&=&\frac{1}{2}\{\omega_0+2\mu^2\mathrm{Im}(\Gamma_++\Gamma_-)\}\sigma_z
\\
&=&\frac{1}{2}\big\{\omega_0+\frac{\mu^2}{2\pi^2}\int^\infty_0d\omega
P(\frac{\omega}{\omega+\omega_0}-\frac{\omega}{\omega-\omega_0})(1+\frac{2}{e^{2\pi\kappa\omega}-1})\big\}
\sigma_z,  \end{eqnarray}
for the atom at rest in the static coordinate system.

Then, the geometric phase for
the atom at rest in the static coordinate system is given by, to the first order,
\begin{eqnarray}\label{GP for static atom}
\Phi_s=-\varphi\sin^2\frac{\theta}{2}-\varphi^2\frac{\gamma_0}{8\omega_0}\sin^2\theta
(2+\cos\theta+\frac{2}{e^{2\pi\kappa\omega_0}-1}\cos\theta).
\end{eqnarray}
Getting rid of the contribution deriving from the inertial atom in Minkowski spacetime, we obtain
\begin{eqnarray}\label{pure geometric phase 2}
\delta\Phi=\Phi_s-\Phi_I=-\varphi^2\frac{\gamma_0}{4\omega_0}\sin^2\theta \frac{1}{e^{2\pi\kappa\omega_0}-1}\cos\theta.
\end{eqnarray}
This is exactly the geometric phase purely induced by the presence of a thermal bath at the temperature
\begin{eqnarray}\label{temperatrue}
T_s=\frac{1}{2\pi\kappa}=\frac{1}{2\pi\alpha\sqrt{g_{00}}},
\end{eqnarray}
felt by the static observer, which is similar to the correction to that of an inertial atom resulting from the existence of a thermal bath at the same temperature in the Minkowski spacetime. It is needed to note that this temperature is different from what was obtained in the case of the freely failing atom $(T_f=1/2\pi\alpha)$, but we can relate
these two temperatures by
\begin{eqnarray}\label{relation between two temperature}
T^2_s=\big(\frac{1}{2\pi\alpha}\big)^2+\big(\frac{a}{2\pi}\big)^2=T^2_f+T^2_U.
\end{eqnarray}
In which the first term is the square of the Gibbons-Hawking temperature of de Sitter spacetime, and the second term is related to the Unruh temperature, which depends on proper acceleration described by Eq. (\ref{proper acceleraion}). Therefore, the correction to the geometric phase of the static atom derives from a combined effect containing both the thermal nature of de Sitter spacetime characterized by the Gibbons-Hawking temperature and the Unruh effect due to the proper acceleration of the atom. Thus, we arrive at the conclusion that the temperature felt by the static atom, in terms of the geometric phase, is the square root of the sum of squared Gibbons-Hawking temperature and the squared Unruh temperature associated with its proper acceleration. It should be noted that the relation in Eq. (\ref{relation between two temperature}) agrees with the result obtained in other different physical contexts \cite{zhou1,Yu,Deser}.

Now, we discuss the detectability of this correction for the static atom from the theory analysis. From Eqs. (\ref{pure geometric phase 2}) and (\ref{relation between two temperature}), the correction corresponding to the static atom depends on not only the Gibbons-Hawking temperature, but also the location of the static atom $r$ (which decides the proper acceleration of the static atom). We rewrite the temperature (\ref{temperatrue}) as
\begin{eqnarray}\label{static temperature}
T_s=\frac{1}{\sqrt{1-\zeta^2}}T_f,
\end{eqnarray}
where $\zeta=\frac{r}{\alpha}$ is constant, which changes from $0$ (corresponding to $r=0$) to $1$
(corresponding to $r=\alpha$). As estimated above, $T_f\approx3.94\times10^{-30}$ K. To get an observable
correction to the geometric phase with $\delta\Phi\sim10^{-5}$, $\frac{1}{\sqrt{1-\zeta^2}}\sim10^{28}$ , for a single period of evolution, is requested if we choose $\frac{\gamma_0}{\omega_0}\sim10^{-6}$ and $\omega_0$ as above, which means that the atom must locate near the horizon. Therefore, an observable magnitude of the correction to geometric phase is obtained only when the atom locates near the horizon. However, in this case the atom, to avoid falling into the horizon, must have the acceleration $a\sim10^{18}m/s^2$, which is a large value. So, the correction to geometric phase is only of theoretical interest. It is also needed to note that in this case, $T_U\gg T_f$, so we have $T_s\approx T_U$, and the correction can be thought to completely come from the Unruh effect.


\section{Conclusions} \label{section 4}

In the framework of open quantum systems, we have studied the nonunitary evolution of both freely falling and static two-level atoms interacting with a conformally coupled massless scalar field in the de Sitter-invariant vacuum, and calculated the geometric phases of these both cases. We find that the geometric phase, for both the freely falling and the static atoms, are in structural similarity to that of an inertial atom immersed in a thermal bath in the Minkowski spacetime.

For the Freely falling atom, the geometric phase gets a correction as if it was immersed in a thermal bath with the Gibbons-Hawking temperature $T_f=1/2\pi\alpha$. This clearly suggests that the intrinsic thermal nature of de Sitter spacetime exists. Furthermore, we find this correction, because of tiny cosmological constant or low Gibbons-Hawking temperature (about $3.94\times10^{-30}~\mathrm{K}$), is very small and is unrealistic for actual experimental measurement.

For the static atom, it is found that the correction to the geometric phase results from a composite effect which contains the Gibbons-Hawking effect of the de Sitter spacetime and the Unruh effect associated with the proper acceleration of the atom. Furthermore, this correction, being similar to the freely falling atom case, just like that of an inertial atom in a thermal bath in the Minkowski spacetime, but the temperature that the static atom feels is a square root of the sum of the squared Gibbons-Hawking temperature and the squared Unruh temperature associated with the atomic proper acceleration. We have also estimated the magnitude of this correction, and found that, since the Gibbons-Hawking temperature is very tiny, by choosing appropriate distance $r$ (corresponding to appropriate proper acceleration), the correction almost completely comes from the Unruh effect.

\begin{acknowledgments}

This work was supported by the  National Natural Science Foundation
of China under Grant No. 11175065, 10935013; the National Basic
Research of China under Grant No. 2010CB833004; the SRFDP under
Grant No. 20114306110003; PCSIRT, No. IRT0964; Hunan Provincial Innovation Foundation For Postgraduate under Grant No CX2012B202; the Hunan Provincial
Natural Science Foundation of China under Grant No 11JJ7001;  and
Construct Program of the National Key Discipline.

\end{acknowledgments}


\end{document}